\documentclass[11pt]{article}  %Do not change the point size.
\usepackage{myprep}
\usepackage{cite}
\usepackage{epsfig}
\usepackage{amsmath,amssymb}
\newcommand{\fract}[2]{{\textstyle\frac{#1}{#2}}}

\newcommand{\fpt}{F^{\prime2}}

\newcommand{\sFt}{{{\rm sin}^2}F}
\newcommand{\lapeq}{\stackrel{\scriptscriptstyle\raisebox{-2.5mm}{$<$}}
{\scriptscriptstyle \raisebox{-0.8mm}{$\sim$}}}

\begin{document}
%
%____________________________________________________________
%
%  Title, authors, institutions, and abstract
%----------------------------------------------------------------
%  Syntax:  \titlematter{title}{authors}{institutions}{abstract}
%----------------------------------------------------------------
%     If lines are too long, use linebreaks where convenient.
%     If all authors are from the same institution, omit raised letters.
%
\titlematter{The Spin of the Nucleon in Effective Models$^*$}%
{H. Weigel$^\dagger$}%
{Institute of Theoretical Physics, T\"ubingen University\\
     Auf der Morgenstelle 14, D--72076 T\"ubingen, Germany}
{The three flavor soliton approach for baryons is utilized 
to discuss effects of flavor symmetry breaking in the baryon
wave--functions on axial current matrix elements. The flavor 
content of the singlet axial current matrix elements,
that parameterizes the quark spin contribution to the total 
angular momentum, is disentangled and studied as a function 
of the effective flavor symmetry breaking. Here the nucleon and the 
$\Lambda$--hyperon are considered.}
%
%
%____________________________________________________________
%  Start article here:

%%%%%%%%%%%%%%%%%%%%%%%%%%%%%%%%%%%%%%%%%%%%%%%%%%%%%%%%%%%%%%%%%%%%%%%%%%%%%%%%%%
\section{Introduction}     
Even though the fundamental theory for the strong interaction processes 
of hadrons, Quantum Chromodynamics (QCD), is well established, hadron 
properties can unfortunately not be computed directly. However, QCD 
contains a hidden expansion parameter, the number ($N_C$) of 
color degrees of freedom, that is beneficial for model building. 
\begin{table}[b]
\noindent
\footnotesize{$^*$\parbox[t]{16cm}{Plenary talk presented at the 
$9^{\rm th}$ International Symposium on Meson-Nucleon Physics and 
the Structure of the Nucleon, Washington DC, July 2001.}}\\
\noindent
\footnotesize{$^\dagger$Heisenberg--Fellow}
\end{table}
For arbitrarily large $N_C$, QCD becomes equivalent to a theory of weakly 
interacting mesons~\cite{tH74}. That is, 
the meson interaction strengths scale like $1/N_C$ 
while baryon masses and radii scale like $N_C$ and $N_C^{0}$, 
respectively~\cite{Wi79}. Meson Lagrangians may possess localized 
solutions to the field equations with finite field energy: solitons. 
Their energies scale inversely with the meson coupling and 
their extensions approach constants as the coupling increases. These 
analogies lead to the conjecture that baryons emerge as solitons 
in the effective meson theory that is equivalent to QCD~\cite{Wi79,Sk61}. 
Although this meson theory cannot be derived from QCD, low--energy meson 
phenomenology provides sufficient constraints to build sensible models. 
Especially chiral symmetry and its breaking in the vacuum introduce 
non--linear interactions for the pions, the (would--be) Goldstone
bosons of chiral symmetry. Then effective Lagrangians are constructed from 
the chiral field 
$U={\rm exp}\left(i\vec{\tau}\cdot\vec{\pi}/f\right)$
that are invariant under global chiral transformations $U\to LUR^\dagger$. 
As $U^\dagger U=1$, at least two derivatives are required
\begin{equation}
{\cal L}_0=\frac{f^2}{4}{\rm tr}
\left(\partial_\mu U \partial^\mu U^\dagger\right)\, .
\label{lag0}
\end{equation}
Extracting the axial current 
$A_\mu^a=f\partial_\mu \pi^a+{\cal O}(\vec{\pi}\hspace{0.2mm}^3)$
from ${\cal L}_0$ provides the electroweak coupling and determines 
the pion decay constant $f=f_\pi=93{\rm MeV}$.
Having established a chiral model, a finite energy soliton 
solution must be obtained and quantized to describe baryon states.
I will outline this approach in section 2. In section 3 I will consider 
three flavor extensions thereof with special emphasis on the role of 
flavor symmetry breaking~\cite{We96}. I will employ these methods to 
compute axial current matrix elements of baryons in section 4. These
matrix elements are major ingredients for the description of the
nucleon spin structure~\cite{Ja96} as they reflect its various quark 
flavor contributions~\cite{El95} and they parameterize hyperon
beta--decay. The effects of flavor symmetry breaking will be essential 
to discuss the strange quark contribution. Section 5 contains 
some concluding remarks.

\section{Baryons as Chiral Solitons}
Scaling considerations show that the model~(\ref{lag0}) does 
not contain stable soliton solutions. Therefore Skyrme added a 
stabilizing term~\cite{Sk61}
\begin{equation}
{\cal L}=\frac{f_\pi^2}{4}{\rm tr} 
\left[\partial_\mu U \partial^\mu U^\dagger\right]
+\frac{1}{32e^2}{\rm tr}\left(
\left[U^\dagger\partial_\mu U,U^\dagger\partial_\nu U\right]
\left[U^\dagger\partial^\mu U,U^\dagger\partial^\nu U\right]\right)\, ,
\label{lagsk}
\end{equation}
that is of fourth order in the derivatives. There are other 
stabilizing extensions of  ${\cal L}_0$, as {\it e.g.} including 
vector mesons~\cite{Ja88,Ka84}. Although such extensions appear 
physically more motivated, I will stick to the Skyrme model for 
pedagogical reasons when explaining the soliton picture for baryons.

The soliton solution to (\ref{lagsk}) assumes the famous hedgehog
shape
\begin{equation}
U_{\rm H}\left(\vec{r}\,\right)=
{\rm exp}\left(i\vec{\tau}\cdot\hat{r}F(r)\right)\,.
\label{hedgehog}
\end{equation}
The equations of motion become an ordinary second order differential
equation for the chiral angle $F(r)$ that is obtained by
extremizing the classical energy 
\begin{equation}
E_{\rm cl}=E_{\rm cl}[F]=\int d^3r \left\{\frac{f_\pi^2}{2}
\left(r^2\fpt+2\sFt\right)+\frac{\sFt}{2e^2}
\left(2\fpt+\frac{\sFt}{r^2}\right)\right\}\, .
\label{ecl}
\end{equation}
It can be argued~\cite{Wi83}
that the baryon number equals the winding number of the  
mapping~(\ref{hedgehog}), {\it i.e.} $B=[F(\infty)-F(0)]/\pi$. Hence
the boundary conditions $F(0)=-\pi$ and $F(\infty)=0$, that correspond
to unit baryon number, determine the chiral angle uniquely. This soliton 
does not yet describe states of good spin and/or flavor as the 
{\it ansatz}~(\ref{hedgehog}) does not possess the corresponding 
symmetries. Such states are generated by restoring these symmetries 
through collective coordinates $A(t)$
\begin{equation}
U(\vec{r},t)\,=\,A(t)\, U_{\rm H}(\vec{r}\,)\,A^\dagger(t)\, .
\label{collrot2}
\end{equation}
and subsequent canonical quantization thereof~\cite{Ad83}. This
introduces right $[A,R_i]=A\tau_i/2$ and left generators 
$[A,L_i]=\tau_i A/2$. While the isospin interpretation $I_i=L_i$ is 
general, the identity $J_i=-R_i$ for the spin is due to the hedgehog 
structure~(\ref{hedgehog}) as is the relation 
$|\vec{I}|=|\vec{J}|$. Quantizing the collective coordinates yields
a Hamiltonian in terms of spin (isospin) operators
\begin{equation}
H_{\rm coll}\,=\,E_{\rm cl}+\frac{\vec{J}\,^2}{2\alpha^2}
\,=\,E_{\rm cl}+\frac{\vec{I}\,^2}{2\alpha^2}\, .
\label{hamrot2}
\end{equation}
The moment of inertia is also a functional of the above
determined chiral angle
\begin{equation}
\alpha^2[F]=\frac{2}{3}\int d^3r\,
{\rm sin}^2F
\left[f_\pi^2+ \frac{1}{e^2}
\left(F^{\prime2}+\frac{{\rm sin}^2F}{r^2}\right)\right]\, .
\label{mominert}
\end{equation}
Matching the mass difference
$M_\Delta-M_{\rm N}=\frac{3}{2\alpha^2}
\sim 300{\rm MeV}$ fixes the undetermined parameter 
$e\approx4.0$.

\section{Extension to Three Flavors}

The generalization to three flavors is carried out straightforwardly
by taking $A(t)\in SU(3)$ with the hedgehog~(\ref{hedgehog}) embedded 
in the isospin subgroup. However, the Lagrangian acquires
two essential extensions. The first one
is the Wess--Zumino--Witten term~\cite{Wi83}. Gauging it for local
$U_V(1)$ shows that indeed the winding number current equals the
baryonic current. Furthermore it constrains $A$ to be quantized as 
a fermion (for $N_C$ odd). 
The second extension originates from flavor symmetry breaking that is
reflected by different masses and decay constants of the pseudoscalar
mesons
\begin{equation}
{\cal L}_{\rm SB}=\frac{f_\pi^2 m_\pi^2 -f_K^2 m_K^2}{2\sqrt3}
{\rm tr}\left\{\lambda_8\left(U+U^\dagger\right)\right\}
%\nonumber \\* &&\hspace{0.2cm}
+\frac{f_K^2-f_\pi^2}{4\sqrt3}{\rm tr}\left\{\lambda_8
\left(\partial_\mu U \partial^\mu U^\dagger U 
+{\rm h.c.}\right)\right\}\, .
\label{lsymbr}
\end{equation}
The explicit form of ${\cal L}_{\rm SB}$ is model dependent, however, 
the techniques to study its effects on baryon properties are general.
The $SU(3)$ collective coordinates are parameterized by 
\underline{eight} ``Euler--angles''
\begin{equation}
A=D_2(\hat{I})\,{\rm e}^{-i\nu\lambda_4}D_2(\hat{R})\,
{\rm e}^{-i(\rho/\sqrt{3})\lambda_8}\ ,
\label{Apara}
\end{equation}
where $D_2$ denote rotation matrices of three Euler--angles for each, 
rotations in isospace~($\hat{I}$) and coordinate--space~($\hat{R}$). 
Substituting the {\it ansatz}~(\ref{collrot2}) into 
${\cal L}+{\cal L}_{\rm SB}$ and canonical quantization of the 
collective coordinates~$A$ yields 
\begin{equation}
H=H_{\rm s}+\fract{3}{4}\, \gamma\, {\rm sin}^2\nu\, .
\label{Hskyrme}
\end{equation}
The symmetric piece of this Hamiltonian only contains Casimir 
operators that may be expressed in terms of the $SU(3)$--right
generators $R_a\, (a=1,\ldots,8)$:
\begin{equation}
H_{\rm s}=E_{\rm cl}+\frac{1}{2\alpha^2}\sum_{i=1}^3 R_i^2
+\frac{1}{2\beta^2}\sum_{\alpha=4}^7 R_\alpha^2\, .
\label{Hsym}
\end{equation}
While $\beta^2$ is a moment of inertia 
similar to $\alpha^2$ in eq~(\ref{mominert}), $\gamma$ 
originates from symmetry breaking
\begin{eqnarray}
\gamma=\gamma[F]=
\frac{2\pi}{3}\int d^3r \left[\left(f_K^2m_K^2-f_\pi^2m_\pi^2\right)
\left(1-{\rm cos}F\right)+\frac{f_K^2-f_\pi^2}{2}
{\rm cos}F\left(F^{\prime2}r^2+2{\rm sin}^2F\right)\right].
\nonumber
\end{eqnarray}
The generators $R_a$ can be expressed in terms
of derivatives with respect to the `Euler--angles'. The eigenvalue
problem $H\Psi=\epsilon\Psi$ reduces to sets of ordinary second order
differential equations for isoscalar functions which only depend on
the strangeness changing angle $\nu$~\cite{Ya88}. Only the product
$\omega^2=\frac{3}{2}\gamma\beta^2$ appears in these differential
equations that are integrated numerically. Thus $\omega^2$ is 
interpreted as the effective strength of the flavor symmetry breaking. 
A value in the range $5\mbox{\scriptsize ${\lapeq}$}\omega^2
\mbox{\scriptsize ${\lapeq}$}8$
is required to obtain reasonable agreement with the empirical mass
differences for the $\frac{1}{2}^+$ and $\frac{3}{2}^+$
baryons~\cite{We96}. The eigenstates of the symmetric piece~(\ref{Hsym})
are members of definite $SU(3)$ representations, {\it e.g.} the
octet ({\bf 8}) for the low--lying $\frac{1}{2}^+$ baryons. Upon flavor 
symmetry breaking, states of different representations are mixed.
At $\omega^2=6$ the nucleon amplitude contains a 23\% contamination
of the state with nucleon quantum numbers in the ${\bf \bar{10}}$ 
representation. This clearly shows a strong deviation
from flavor covariant wave--functions.

\section{Axial Current Matrix Elements}

\begin{table}[b]
~\vskip-0.5cm
\caption{\label{empirical}\sf The empirical values for the
$g_A/g_V$ ratios of hyperon beta--decays \protect\cite{DATA}.
For $\Sigma\to\Lambda$ only $g_A$ is given.
Also the flavor symmetric predictions are presented using the
values for $F$\&$D$ of Ref.~\protect\cite{Fl98}.}
~\vskip0.01cm
\centerline{\small
\begin{tabular}{ c || c| c | c | c | c | c}
& $n\to p$ & $\Lambda\to p$ & $\Sigma\to n$ & $\Xi\to\Lambda$ &
$\Xi\to\Sigma$ & $\Sigma\to\Lambda$\\
\hline
emp.&$1.267\pm0.004$& $0.718\pm0.015$ & $0.340\pm0.017$ 
& $0.25\pm0.05$ & $1.287\pm0.158$ & $0.61\pm 0.02$\\
$F$\&$D$& $1.26=g_A$ & $0.725\pm0.009$ & $0.339\pm0.026$ 
& $0.19\pm0.02$ & $1.26=g_A$ & $0.65\pm0.01$
\end{tabular}}
\end{table}

The effect of the derivative type symmetry
breaking terms is mainly indirect. They provide the splitting between the
various decay constants and thus increase $\gamma$ because of
$f_K^2m_K^2-f_\pi^2m_\pi^2\approx 1.5f_\pi^2(m_K^2-m_\pi^2)$.
Otherwise the ($f_K^2-f_\pi^2$)--terms may be omitted. Whence there 
are no symmetry breaking terms in current operators and the non--singlet 
axial charge operator is parameterized as 
\begin{equation}
\int d^3r A_i^{(a)} = c_1 D_{ai} - c_2 D_{a8}R_i
+c_3\sum_{\alpha,\beta=4}^7d_{i\alpha\beta}D_{a\alpha}R_\beta
\, ,\quad c_i=c_i[F] \, ,
\label{axsym}
\end{equation}
where
$D_{ab}=\frac{1}{2}{\rm tr}\left(\lambda_a A\lambda_b A^\dagger\right)$,
$ a=1,\ldots,8$ and $i=1,2,3$.
When integrating out {\it strange} degrees of freedom, 
$\omega^2\to\infty\, $ the strangeness contribution to the nucleon
axial charge should vanish. The eigenstates of~(\ref{Hskyrme})
parametrically depend on $\omega^2$ and for
$\omega^2\to\infty$ the singlet current
\begin{equation}
\int d^3r A_i^{(0)}= -2\sqrt{3} c_2 R_i\, , \quad (i=1,2,3)
\label{singsym}
\end{equation}
yields a vanishing nucleon matrix element of the  strangeness
projection, $A_i^{(s)}=(A_i^{(0)}-2\sqrt{3}A_i^{(8)})/3$.
The identity of $c_2$ in eqs~(\ref{axsym}) 
and~(\ref{singsym}) goes beyond group theoretical arguments.
Actually all model calculations in the literature \cite{Pa92,Bl93}
are consistent with (\ref{singsym}).
To completely describe the hyperon beta--decays I 
demand matrix elements of the vector charges that are obtained
from the operator
\begin{equation}
\int d^3r V_0^{(a)} = \sum_{b=1}^8D_{ab}R_b=L_a\, .
\label{vector}
\end{equation}

\begin{figure}[t]
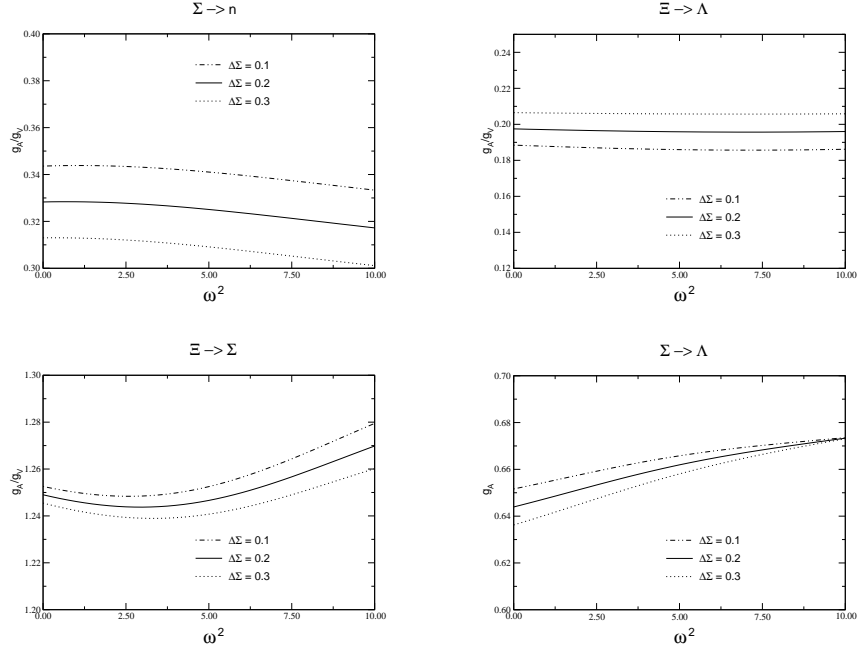

~\vskip0.05cm
\centerline{
\epsfig{figure=nusi.eps,height=5.0cm,width=4.0cm,angle=270}
\hspace{1.0cm}
\epsfig{figure=chla.eps,height=5.0cm,width=4.0cm,angle=270}}
~\vskip0.03cm
\centerline{
\epsfig{figure=sich.eps,height=5.0cm,width=4.0cm,angle=270}
\hspace{1.0cm}
\epsfig{figure=sila.eps,height=5.0cm,width=4.0cm,angle=270}}
\caption{\label{decay}\sf The predicted decay parameters for the
hyperon beta--decays using $\omega^2_{\rm fix}=6.0$.
The errors originating from those in $\Delta\Sigma_N$ are indicated.}
\end{figure}

The values for $g_A$ and $g_V$
(only $g_A$ for $\Sigma^+\to\Lambda e^+\nu_e$)
are obtained from the matrix elements of respectively
the operators in eqs~(\ref{axsym}) and~(\ref{vector}), sandwiched
between the eigenstates of the full Hamiltonian~(\ref{Hskyrme}).
I choose $c_2$ according to
$\langle N \uparrow |\int d^3r A_3^{(0)} |N\uparrow
\rangle=\sqrt3 c_2= \Delta\Sigma=0.2\pm0.1$~\cite{El95} and 
subsequently determine $c_1$ and $c_3$ at $\omega^2_{\rm fix}=6.0$ 
such that the empirical values for the nucleon axial charge, $g_A$ 
and the $g_A/g_V$ ratio for $\Lambda\to p e^-\bar{\nu}_e$ are 
reproduced\footnote{Here the problem of the too small model prediction 
for $g_A$ will not be addressed but rather the empirical value 
$g_A=1.26$ will be used as an input to fix the $c_n$.}. This 
predicts the other decay parameters and describes their variation 
with symmetry breaking as shown in figure~\ref{decay}.
The dependence on flavor symmetry breaking is very
moderate\footnote{However, the individual matrix elements
entering the ratios $g_A/g_V$ vary strongly with
$\omega^2$~\cite{We00}.} and the results can be viewed as reasonably
agreeing with the empirical data, {\it cf.} table \ref{empirical}.
The two transitions, $n\to p$ and $\Lambda\to p$, which are not shown in
figure~\ref{decay}, exhibit a similar negligible dependence on $\omega^2$.
Hence these predictions
are not sensitive to the choice of $\omega^2_{\rm fix}$.
Comparing the results in figure \ref{decay} with the data in
table~\ref{empirical} shows that the calculation using the strongly
distorted wave--functions agrees equally well with the empirical
data as the established~\cite{Fl98} flavor symmetric $F$\&$D$ fit. 

\begin{figure}[t]
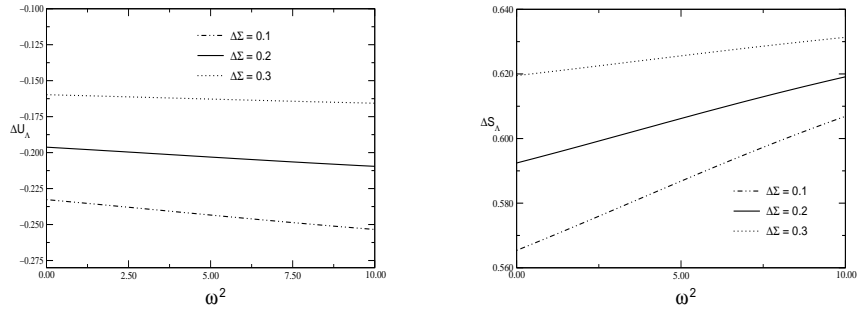

~\vskip-0.03cm
\centerline{
\epsfig{figure=h1la.eps,height=5.0cm,width=4.0cm,angle=270}
\hspace{1.0cm}
\epsfig{figure=h3la.eps,height=5.0cm,width=4.0cm,angle=270}}
\caption{\label{laxial}\sf The contributions of the {\it non--strange}
(left panel) and {\it strange} (right panel) degrees of freedom
to the axial charge of the $\Lambda$. Again $\omega^2_{\rm fix}=6.0$ 
was assumed.}
\end{figure}

Figure \ref{laxial} shows the flavor components of the axial
charge of the $\Lambda$ hyperon. Again, the various contributions
to the axial charge of the $\Lambda$ exhibit only moderate dependences
on $\omega^2$. The {\it non--strange} component,
$\Delta U_\Lambda=\Delta D_\Lambda$ slightly increases in magnitude.
The {\it strange} quark piece, $\Delta S_\Lambda$ grows with
symmetry breaking since $\Delta\Sigma_\Lambda$ is kept fixed. These
results nicely agree with an $SU(3)$ analysis applied to the
data~\cite{Ja96a}. 

The observed independence on $\omega^2$ does not 
occur for all matrix elements of the axial current. A prominent exemption
is the {\it strange} quark component in the nucleon, $\Delta S_N$.
For $\Delta\Sigma=0.2$, say, it is significant at zero symmetry
breaking, $\Delta S_N=-0.131$ while it decreases (in magnitude) to
$\Delta S_N=-0.085$ at $\omega^2=6.0$.

This far I have only considered the general sturcture of the
current operators without computing
the constants $c_i$ from a model soliton, though I had the Skyrme model 
in mind. However, this model is too simple to be realistic. For 
example, it improperly predicts $\Delta\Sigma=0$~\cite{We96}. 
More complicted models must be utilized, as {\it e.g.}
the vector meson model that has been established for 
two flavors in ref~\cite{Ja88}. Later it has been generalized to 
three flavors and been shown to fairly describe hyperon
beta--decay~\cite{Pa92}. To account for different masses and decay 
constants a minimal set of symmetry breaking terms is 
included~\cite{Ja89} that add
symmetry breaking pieces to the axial charge operator,
$$
\delta A_i^{(a)}=c_4 D_{a8}D_{8i}+
c_5\hspace{-1.0mm} \sum_{\alpha,\beta=4}^7
d_{i\alpha\beta}D_{a\alpha}D_{8\beta}+
c_6 D_{ai}(D_{88}-1)\,\, ,\,
\delta A_i^{(0)}= 2\sqrt{3}\,c_4D_{8i}\, .
$$
The coefficients $c_1,\ldots,c_6$ are functionals of the soliton 
and can be computed once the soliton is constructed~\cite{We00}.
As the model parameters cannot be completely determined in the 
meson sector~\cite{Ja88} I use the small remaining freedom to 
accommodate baryon properties in three different ways, see
table \ref{realistic}. The set denoted by `masses' refers 
to a best fit to the baryon mass differences.
It predicts the axial charge somewhat on the low side, $g_A=0.88$.
The set named `mag.mom.' refers to parameters that yield 
magnetic moments of the $\frac{1}{2}^+$ baryons close to the 
respective empirical data (with $g_A=0.98$) and finally the set 
labeled `$g_A$' reproduces~\cite{Pa92} the axial charge of the nucleon 
as well as the hyperon beta--decay data.
\begin{table}[t]
\caption{\label{realistic}\sf Spin content of the $\Lambda$ in the
realistic vector meson model. For comparison the nucleon
results are also given. Three sets of model parameters
are considered, see text.}
~\vskip0.01cm
\centerline{\small
\begin{tabular}{ c || c | c |c || c | c | c | c}
& \multicolumn{3}{c||}{$\Lambda$} &
\multicolumn{4}{c}{$N$}\\
\hline
``fits'' & $\Delta U = \Delta D$ & $\Delta S$ & $\Delta\Sigma$ &
 $\Delta U$ & $\Delta D$ & $\Delta S$ & $\Delta\Sigma$\\
\hline
masses
&$-0.155$&$0.567$&$0.256$&$0.603$&$-0.279$&$-0.034$&$0.291$\\
mag. mom.
&$-0.166$&$0.570$&$0.238$&$0.636$&$-0.341$&$-0.030$&$0.265$\\
$g_A$
&$-0.164$&$0.562$&$0.233$&$0.748$&$-0.476$&$-0.016$&$0.256$
\end{tabular}}
\end{table}
As presented in table~\ref{realistic}, the predictions for the axial 
properties of the $\Lambda$ are insensitive to the model parameters.
The singlet matrix element of the $\Lambda$ hyperon is 
smaller than that of the nucleon. Sizable polarizations of the {\it up} 
and {\it down} quarks in the $\Lambda$ are again predicted. They are 
slightly smaller in magnitude but nevertheless comparable to those 
obtained from the $SU(3)$ symmetric analyses~\cite{Ja96a}.

\section{Conclusions}
In this talk I utilized the picture that baryons emerge as solitons
in an effective meson theory to compute various baryon matrix elements. 
Here I focused on the effects of flavor symmetry breaking in the baryon
wave--functions and showed that despite of strong deviations
from flavor covariant wave--functions the empirical parameters
for hyperon beta--decay are reproduced. Effective symmetry 
breaking is treated as a parameter and consistency with the 
the two--flavor limit (infinitely heavy strange quarks) relates
singlet and octet axial currents beyond group theory. With this I
showed that chiral soliton models explain the proton spin puzzle, 
{\it i.e.} the smallness of the observed
axial singlet current matrix element. Furthermore flavor symmetry breaking 
in the nucleon wave--function significantly reduces the polarization of 
the strange quarks inside the nucleon.

\acknowledgments{
This work has been supported by the Deutsche Forschungsgemeinschaft
under contracts We 1254/3-1 and We 1254/4-2.}

%%%%%%%%%%%%%%%%%%%%%%%%%%%%%%%%%%%%%%%%%%%%%%%%%%%%%%%%%%%%%%%%%%%%%%%%%%%%%%%%%%
%____________________________________________________________
%  Start references here:

\end{document}